\documentclass[reprint,
superscriptaddress,
amsmath,amssymb,
notitlepage,
aps,
prl,
10pt, 
tightenlines
]{revtex4-2}

\usepackage{graphicx}
\usepackage{amsmath}
\usepackage{amsthm}
\usepackage{amsfonts}
\usepackage{dsfont}
\usepackage[normalem]{ulem}
\usepackage{lipsum}
\usepackage{xcolor}
\usepackage{tikz}
\usepackage{subcaption}

\usetikzlibrary {decorations.pathmorphing, decorations.pathreplacing, decorations.shapes} 

\DeclareMathOperator{\tr}{tr}
\DeclareMathOperator{\pf}{pf}

\DeclareMathOperator{\ad}{ad}
\newcommand{\ket}[1]{|#1\rangle}
\newcommand{\bra}[1]{\langle#1|}
\newcommand{\ev}[1]{\langle#1\rangle}
\newcommand{\ketbra}[2]{|#1\rangle\hspace{-.6mm}\langle#2|}
\newcommand{\mc}[1]{\mathcal{#1}}
\newcommand{\mb}[1]{\mathbf{#1}}
\newcommand{\Z}{\mathbb{Z}}

\newcommand{\R}{\mathbb{R}}
\newcommand{\C}{\mathbb{C}}
\newcommand{\CP}{\mathbb{CP}}
\newcommand{\hilb}{\mathcal{H}}
\newcommand{\ha}{\frac{1}{2}}

\newcommand{\jfrac}[2]{\genfrac[]{0pt}{0}{#1}{#2}}

\begin{document}
\title{Matrix product states as thin torus limits of conformal correlators}
\author{Adrián Franco-Rubio}
\affiliation{Max-Planck-Institut f\"ur Quantenoptik, Hans-Kopfermann-Str.~1, D-85748~Garching, Germany}%
\affiliation{Munich Center for Quantum Science and Technology (MCQST), Schellingstr.~4, D-80799~Munich, Germany}%
\affiliation{University of Vienna, Faculty of Physics, Boltzmanngasse 5, 1090 Vienna, Austria}%
\author{J. Ignacio Cirac}
\affiliation{Max-Planck-Institut f\"ur Quantenoptik, Hans-Kopfermann-Str.~1, D-85748~Garching, Germany}%
\affiliation{Munich Center for Quantum Science and Technology (MCQST), Schellingstr.~4, D-80799~Munich, Germany}%
\author{Germán Sierra}
\affiliation{Instituto de Física Teórica, UAM-CSIC, Universidad Autónoma de Madrid, Cantoblanco, Madrid, Spain}%
\affiliation{Kavli Institute for Theoretical Physics, University of California, Santa Barbara, CA 93106, USA}%
\date{\today}

\begin{abstract}
    We introduce one-parameter families of spin chain ansatz wavefunctions constructed from chiral conformal field theory correlators on a torus, with the modular parameter $\tau$ serving as the deformation parameter. In the cylinder limit $\tau\to\infty$, these wavefunctions reduce to infinite dimensional matrix product states \cite{CiracSierra10}. In contrast, in the thin torus limit $\tau\to 0$, they become finite bond-dimension matrix product states (MPS). Focusing on families derived from the SU(2)$_1$ and SU(2)$_2$ Wess-Zumino-Witten models, we show that in the thin torus limit they reproduce known MPS ground states, such as those of the Majumdar-Ghosh and AKLT spin chains.
\end{abstract}

\maketitle

In the study of strongly correlated quantum many-body systems, families of ansatz wavefunctions are a useful tool. Even if the quantum state of the system being analyzed cannot be exactly represented as a member of the family, it may nevertheless display similar physics: in particular, they may belong to the same phase of matter. A well parameterized family of ansatz wavefunctions can also serve as a variational manifold to approximate the ground state of a given target Hamiltonian.

Tensor networks have been used profusely in the last two decades to give rise to such families of ansatz wavefunctions. Among these, in one spatial dimension, matrix product states (MPS) have proved to be extremely useful in efficiently approximating ground states of both gapped and gapless local Hamiltonians \cite{RevModPhys.93.045003}. Another very important corpus of ansatz wavefunctions arose from the study of the fractional quantum Hall effect \cite{Laughlin} and was soon imported from electron to spin systems, including chiral spin liquids \cite{KalmeyerLaughlin}. These wavefunctions turn out to be intimately connected to conformal field theory (CFT) correlators, thus establishing a bulk-boundary correspondence with the effective theory of gapless edge modes \cite{MooreRead}.

Combining features of both of these ansatz families, Ref.~\cite{CiracSierra10} introduced infinite-dimensional matrix product states (idMPS) \footnote{In \cite{CiracSierra10} and many subsequent publications, this ansatz was called infinite MPS (iMPS). To avoid confusion with the unrelated ansatz of finite dimensional MPS on an infinite chain, we slightly modify the nomenclature, following a proposal by H.-H. Tu.}, which extend the standard MPS framework by replacing the finite-dimensional virtual space with an infinite-dimensional Hilbert space corresponding 
to a CFT. The amplitudes of the ansatz wavefunction are thus calculated from CFT correlators on the plane, or equivalently on the cylinder. This framework enables the representation of many-body states that lie beyond the scope of finite bond-dimension MPS, including states with infinite correlation length, such as the ground and excited states of CFTs.

It is important to study which of the many useful features and structures known from finite bond-dimensional tensor networks persist and can be harnessed upon this generalization of the ansatz. It has been found that these new states can be recast as the result of sewing generalized tensors, giving rise to so-called field tensor network states \cite{Nielsen_2021}. In this Letter, we adopt a different approach, and ask ourselves if there are quantum field theories for which the resulting idMPS are actually MPS. 

\begin{figure}
    \centering
    \includegraphics[width=\linewidth]{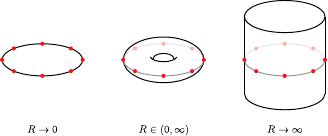}
    \caption{Our one-parameter family of wavefunctions is derived from conformal blocks defined on a torus with modular parameter $\tau\equiv iR$. The \textit{thin torus limit} $R\to 0$ gives rise to a finite bond dimension MPS while the \textit{cylinder limit} $R\to\infty$ recovers the infinite dimensional MPS (idMPS) from \cite{CiracSierra10}.}
    \label{fig:taus}
\end{figure}

Interestingly, we find that MPS naturally emerge when constructing idMPS from CFT correlators with operator insertions along one cycle of a torus, in the limit where the radius $R$ of the orthogonal cycle tends to zero. We refer to this procedure as the \textit{thin torus limit}, borrowing the term from its use in the study of fractional quantum Hall systems \cite{PhysRevLett.94.026802} and fractional topological insulators \cite{bernevig2012}. In this context, the limit corresponds to a geometric deformation of the two-dimensional system into a quasi-one-dimensional configuration, effectively transforming the torus into a narrow cylinder or strip, depending on the boundary conditions. Our construction then places the resulting MPS at one end of a one-parameter family of wavefunctions, indexed by the modular parameter of the torus, depicted in Fig.~\ref{fig:taus}. At the other end, when $R\to\infty$, the torus becomes a cylinder and we recover the known idMPS from the literature. The resulting family thus interpolates between finite and infinite correlation length states.

The paper starts by reviewing the idMPS ansatz and showing how placing it on a torus can give MPS with finite bond dimension. We then look at wavefunctions built from this idea, using SU(2)$_1$ and SU(2)$_2$ WZW models as virtual CFTs. These lead to spin-$\ha$ and spin-1 chains whose thin torus limits match the known MPS for the Majumdar-Ghosh and AKLT models. We end with a discussion, and technical details are in the Supplemental Material.

\textit{Infinite dimensional matrix product states} --- An idMPS is an ansatz wavefunction $\ket{\Psi}$ for a spin system of size $N$ given by
\begin{equation}
    \Psi_{s_1,s_2\ldots,s_N} \equiv \ev{\mc \phi^{s_1}(z_1)\mc \phi^{s_2}(z_2)\ldots\mc \phi^{s_N}(z_N)}_{\text{CB}},
    \label{eq:defidMPS}
\end{equation}
where the LHS is the amplitude of the (unnormalized) wavefunction in a particular $N$ spin configuration and the RHS is the evaluation of a \textit{conformal block} in a given chiral CFT. Recall that conformal blocks are the building blocks of the $N$-point functions of primary fields, and are labeled by the possible fusion paths of the $N$ primary fields. Traditionally, the CFT is defined on the complex plane, or equivalently on the cylinder (since both are related to each other by a conformal transformation that only contributes an overall factor to the wavefunction), but it can be defined on any Riemann surface. The positions of the operator insertions, $z_i$, can be thought of as potential variational parameters. In this work, we will restrict ourselves to 1d translationally invariant spin chains, and we will accordingly fix the insertions to lie equidistantly along a circle, see Fig.~\ref{fig:taus}.

The SU(2)$_k$ Wess-Zumino-Witten (WZW) CFTs \cite{dMFS} constitute specially suitable choices for the virtual CFT when building idMPS. Their primary fields naturally fall into representations of SU(2), making the identification with spins very natural. Crucially, the ones we will use also have free field representations, which allow for easy computation of the conformal blocks. Because of these properties, idMPS for these CFTs have already been extensively studied \cite{CiracSierra10,Nielsen_2011}.

\textit{Motivating example: AKLT state from correlators} --- The following example motivates the general construction presented in this paper. Here we show how to obtain the AKLT state \cite{PhysRevLett.59.799}, a well-known exact MPS ground state of a spin-1 local Hamiltonian, from the correlation functions of a virtual theory. In a particular local basis, an MPS tensor $A^a$ representating this state is given by the Pauli matrices, $A^a = \sigma_a$, $a=1,2,3$, which satisfy the $n=3$ Clifford algebra $\{\sigma_a,\sigma_b\}=2\delta_{ab}$. We will consider a one-dimensional QFT, i.e. a quantum mechanical system, resulting from quantizing the following action of three classical Grassmann variables $\chi_a, a=1,2,3$: 
\begin{equation}
    S = -\int{dt\;i\chi_a\partial_t\chi_a}.
    \label{eq:Lag}
\end{equation}
This is a fully constrained theory: due to the lack of a potential term, the Hamiltonian vanishes, as the canonical momenta are proportional to their conjugate variables,
\begin{equation}
    \pi_a = i\chi_a\equiv i q_a\implies L =\dot q_a\pi_a  \implies H = 0.
\end{equation}
Meanwhile, the canonical quantization of the theory leads to the anticommutation relations of the Clifford algebra up to a proportionality factor, $\{\hat\chi_a,\hat\chi_b\}=\delta_{ab}$. We then define a candidate wavefunction for a spin 1 system,
\begin{equation}
    \psi_{a_1\ldots a_N} \equiv \tr{\left(\hat\chi_{a_1}(t_1)\ldots\hat\chi_{a_N}(t_N)\right)}.
    \label{eq:1didMPS}
\end{equation}
Since time evolution is trivial due to the vanishing Hamiltonian, these amplitudes only depend on the ordering of $t_1,\ldots,t_N$ and not on their specific values, and the trace can be interpreted as the expectation value in a thermal state of arbitrary finite $\beta$ (for all these reasons this setting is sometimes referred to as \textit{topological} quantum mechanics \cite{PhysRevD.41.661}). From the fact that the relevant Clifford algebra has a 2-dimensional irrep given by the Pauli matrices $\sigma_a$, it follows that \eqref{eq:1didMPS} describes an MPS wavefunction with tensor $A^a = \sigma_a$, giving rise to the AKLT ground state.

The crucial observation leading to the construction presented in this work is that the action \eqref{eq:Lag} is related to the following action of three Majorana fermions in 2d, which is conformal,
\begin{equation}
    S = -\int{dt\,dx\;i\chi_a(\partial_t+\partial_x)\chi_a}.
    \label{eq:Lag2d}
\end{equation}
Intuitively, \eqref{eq:Lag} can be interpreted as arising from \eqref{eq:Lag2d} in a limit when the second dimension disappears. A natural way to implement this is by defining the two-dimensional theory on a torus, and taking a limit where the radius of the cycle perpendicular to the line of field insertions goes to zero, effecting a dimensional reduction. This would put the AKLT state in one end of a one-parameter family of wavefunctions defined by a free fermion CFT in tori with different radii, as depicted in Fig.~\ref{fig:taus}. 

More generally, defining idMPS from conformal blocks on the torus gives rise to families of spin wavefunctions indexed by the modular parameter $\tau$ of the torus \footnote{A torus of modular parameter $\tau$ can be seen as arising from quotienting the complex plane by the relations $z\sim z+1$ and $z\sim z + \tau$. This completely parameterizes the moduli space of complex structures on the torus.}. We will occupy ourselves with tori for which $\tau = iR$, $R>0$, and pick coordinates such that the two radii are 1 and $R$. Fixing the operator insertions along the real axis of the torus then amounts to $z_i = i/N$. The limit $R\to \infty$ results in the cylinder idMPS construction, while $R\to 0$, the thin torus limit, is where an exact finite dimensional MPS state will arise. As it is argued in \cite{SM}, the fact that states with finite bond dimension arise from thin torus limits is not a coincidence, and can be interpreted as resulting from the suppression of all descendant states in the Hilbert space of the virtual CFT as $\tau\to 0$.

\textit{idMPS from the SU(2)$_1$ WZW model} --- The chiral SU(2)$_1$ WZW model has central charge $c=1$, and three primary fields, a spin singlet and a doublet, at the basis of its two conformal towers, $\mathbf{0}$ and $\mathbf{\frac{1}{2}}$. The only nontrivial fusion rule is $\mathbf{\frac{1}{2}} \times \mathbf{\frac{1}{2}} = \mathbf{0}$.

\begin{figure}
    \centering
        \begin{tikzpicture}
        \def\dx{3}
        \def\dy{0.7}
        \def\sep{2}
            \tikzset{c/.style={insert path={circle[radius=2pt]}}}
            \filldraw (0,\dy) [c] node[left=2] {$\ket{\text{MG}}_\pm$} -- ++(\dx,0) [c] node[right=2] {$\ket{\text{HS}^*}$};
            \filldraw (0,0) [c] node[left=2] {$\ket{\text{MG}}_\mp$} -- ++(\dx,0) [c] node[right=2] {$\ket{\text{HS}}$};
            \draw  (0, \dy+.7) node[] {$\tau = 0$} ++ (\dx,0) node[] {$\tau = i\infty$};
            \node[anchor=south] at (\dx/2, \dy-0.05) {$\small{\psi_\ha}$};
            \node[anchor=south] at (\dx/2, -0.05) {$\small{\psi_0}$};
        \end{tikzpicture}
    \caption{Schematic representation of the wavefunctions arising from the SU(2)$_1$ WZW model.}
    \label{fig:schema1}
\end{figure}
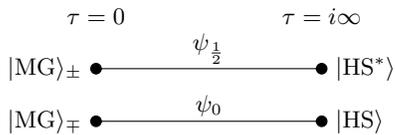

\begin{figure*}

\centering
\begin{subfigure}{0.45\textwidth}
    \includegraphics[width=\textwidth]{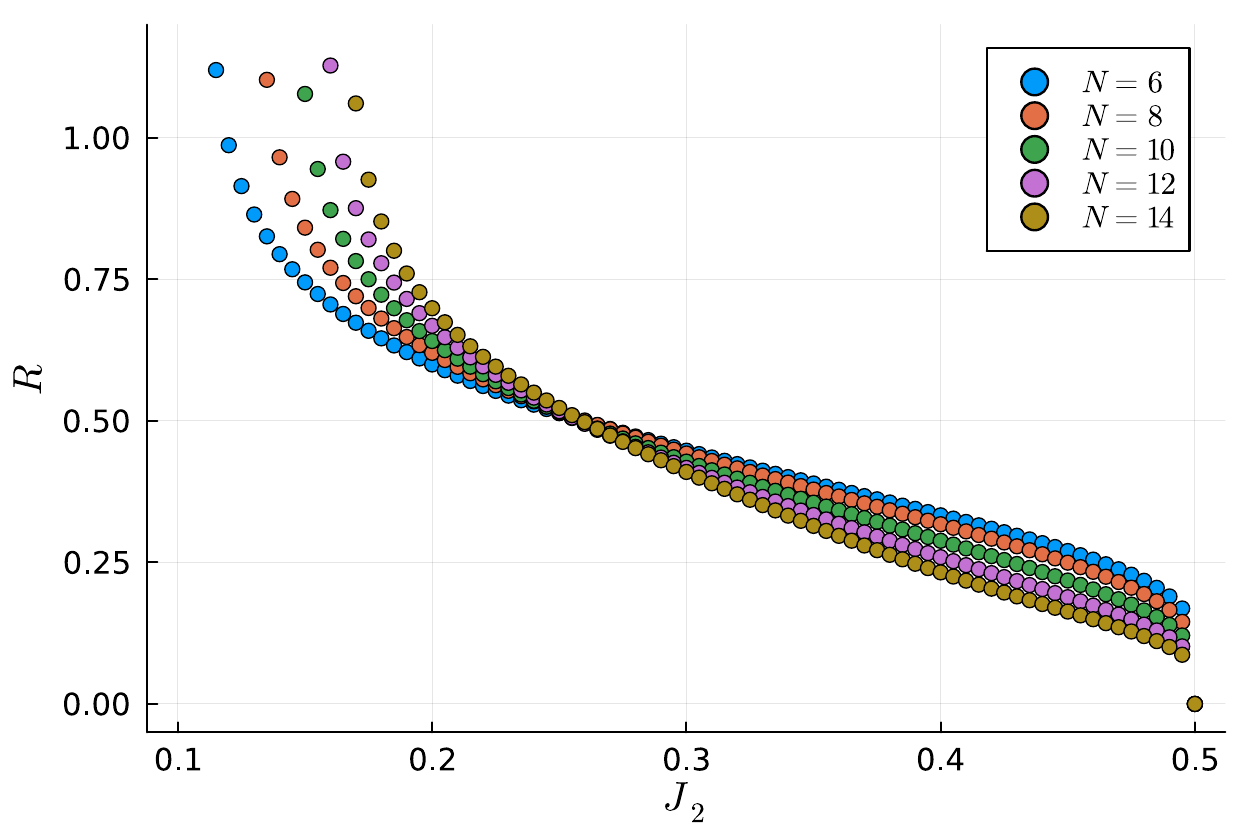}
    \caption{}
    \label{fig:first}
\end{subfigure}
\hfill
\begin{subfigure}{0.45\textwidth}
    \includegraphics[width=\textwidth]{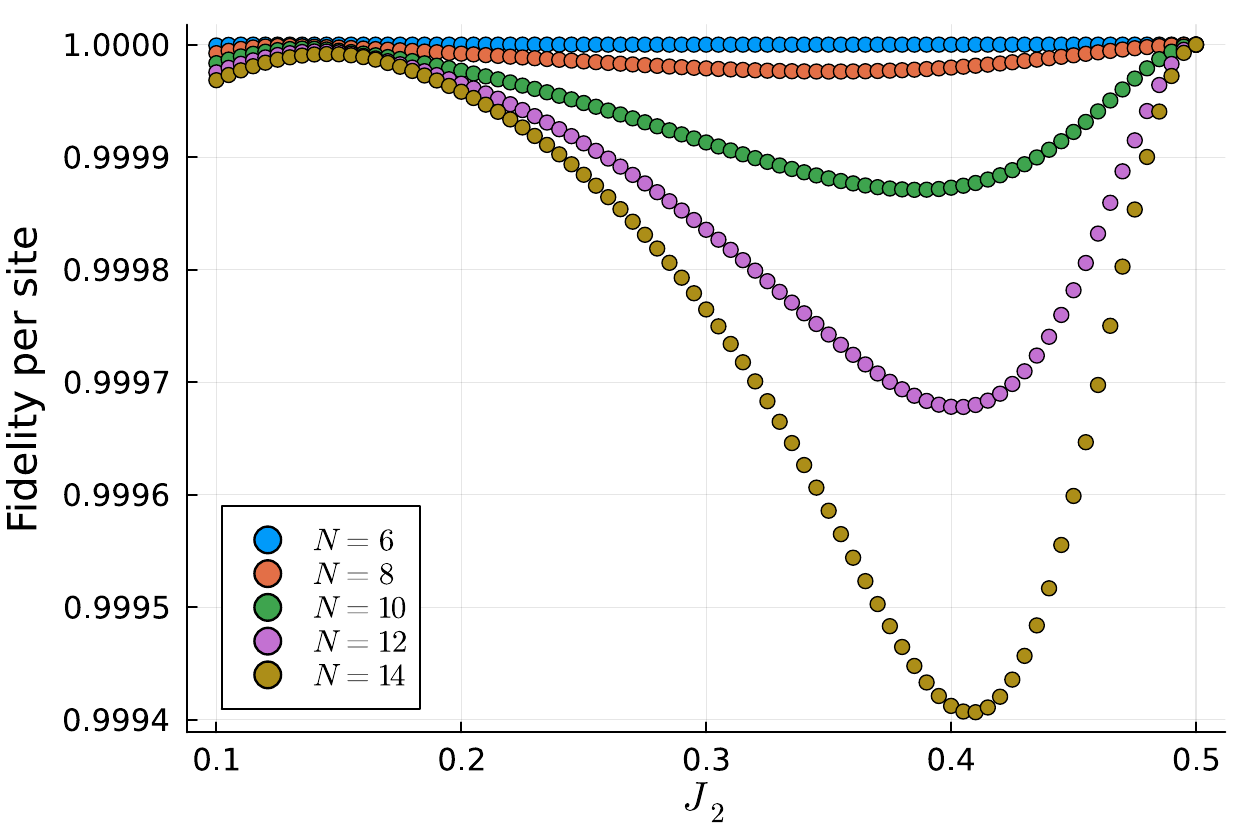}
    \caption{}
    \label{fig:second}
\end{subfigure}
\caption{(a) Value of $R$ that minimizes the energy of $\ket{\psi_0}$ for each $J_2$. Note that for $J_2=0.5$, the ansatz is exact for $R=0$. (b) Fidelity per site of $\ket{\psi_0}$ with the ground state at the optimal value of $R$.}
\label{fig:plots1}
\end{figure*}

We consider the idMPS defined from $N$ insertions of primary fields $\phi_{\frac{1}{2}}^s(z), s=\pm1$, thus giving rise to a wavefunction for a spin $\frac{1}{2}$ chain. We need to ensure that $N$ is even in order for the wavefunction to not vanish. To compute the conformal blocks, we resort to the \textit{free field representation} of the SU(2)$_1$ WZW model, which allows us to express them in terms of correlation functions of a compactified chiral free boson $\varphi(z)$ at radius $r=\sqrt{2}$, according to
\begin{equation}
    \phi^s_{\ha}(z)\longmapsto \mc V_s(z)=e^{is\frac{\varphi(z)}{\sqrt{2}}}.
    \label{eq:mapping}
\end{equation}
The space of conformal blocks for $N$ insertions of $\phi_{\ha}$ on a torus is two-dimensional, corresponding to the two allowed labellings of the fusion graph
\begin{equation}
    \begin{tikzpicture}
    \def\dx{0.5}
    \def\dxb{0.7}
    \def\ddx{0.2}
    \def\ddxb{0.4}
    \def\dy{0.4}
    \def\ddy{0.3}
    \def\dyb{0.65}
        \draw (-\ddx, 0) -- (2*\dx+\dxb+\ddxb, 0) -- (2*\dx+\dxb+\ddxb, -\ddy) -- (-\ddx, -\ddy) -- cycle;
        \foreach \x in {0,\dx,2*\dx, 2*\dx+\dxb}{
        \draw (\x,0) --++ (0, \dy);
        \node[] at (\x, \dyb) {$_\ha$};}
        \node[anchor = south] at (2*\dx+0.5*\dxb, -0.05) {$_{\ldots}$};
        \node[anchor = south] at (0.5*\dx, -0.05) {$_\ha$};
        \node[anchor = south] at (1.5*\dx, -0.05) {$_0$};
        \node[anchor = south] at (2*\dx+\dxb+0.5*\ddxb, -0.05) {$_0$};
    \end{tikzpicture}
    ~,\qquad
    \begin{tikzpicture}
    \def\dx{0.5}
    \def\dxb{0.7}
    \def\ddx{0.2}
    \def\ddxb{0.4}
    \def\dy{0.4}
    \def\ddy{0.3}
    \def\dyb{0.65}
        \draw (-\ddx, 0) -- (2*\dx+\dxb+\ddxb, 0) -- (2*\dx+\dxb+\ddxb, -\ddy) -- (-\ddx, -\ddy) -- cycle;
        \foreach \x in {0,\dx,2*\dx, 2*\dx+\dxb}{
        \draw (\x,0) --++ (0, \dy);
        \node[] at (\x, \dyb) {$_\ha$};}
        \node[anchor = south] at (2*\dx+0.5*\dxb, -0.05) {$_{\ldots}$};
        \node[anchor = south] at (0.5*\dx, -0.05) {$_0$};
        \node[anchor = south] at (1.5*\dx, -0.05) {$_\ha$};
        \node[anchor = south] at (2*\dx+\dxb+0.5*\ddxb, -0.05) {$_\ha$};
    \end{tikzpicture}
    ~.
    \label{eq:trees1}
\end{equation}
We pick a basis for this space labeled by $k=0,\ha$, depending on the operator that ``closes the torus'' in the direction perpendicular to the line of insertions (we discuss this choice in \cite{SM}), and use \eqref{eq:defidMPS} to obtain two families of idMPS wavefunctions, given explicitly by \cite{Dijkgraaf:1987vp, Nielsen_2014}
\begin{align}
    \psi_k(\mb s|\tau) &= \delta_{\mb s}\,\eta_{\mb s}\prod_{i<j}{E(z_i-z_j|\tau)^{\frac{s_is_j}{2}}}\theta\jfrac{k}{0}\left(\sum_j{s_jz_j}\left|2\tau\right.\right),
    \label{eq:wfs1}
\end{align}
where the Jacobi $\theta$ functions and the prime form $E$ are special functions (see \cite{SM}), $\delta_{\mb s}$ is a Kronecker delta enforcing charge neutrality $\sum_i{s_i}=0$, and, following \cite{CiracSierra10}, we have added an additional phase factor $\eta_{\mathbf{s}}$, the \textit{Marshall sign}, defined by 
\begin{equation}
    \eta_{\mathbf{s}} = \prod_{i=1}^{N/2}{s_{2i-1}}\in \{+1,-1\},
    \label{eq:MS}
\end{equation}
corresponding to a local unitary transformation, which ensures the resulting conformal blocks will are singlets under SU(2), as required by the Ward identities \cite{SM}.

\noindent \textit{Cylinder limit} --- It was shown in \cite{CiracSierra10} that the cylinder limit of $\psi_0$, $\lim_{\tau\to i\infty}{\psi_{0}(\mb s|\tau)}$, is the ground state $\ket{\text{HS}}$ of the Haldane-Shastry Hamiltonian \cite{PhysRevLett.60.635, PhysRevLett.60.639},
\begin{equation}
    H_{\text{HS}} = \sum_{i<j}{\frac{\vec S_i\cdot\vec S_{j}}{\sin^2(\pi(i-j)/N)}},
    \label{eq:HS_Ham}
\end{equation}
a gapless, all-to-all generalization of the Heisenberg antiferromagnet. Furthermore, we show in \cite{SM} that the limit of the other wavefunction, $\lim_{\tau\to i\infty}{\psi_{\ha}(\mb s|\tau)}$ is an excited eigenstate $\ket{\text{HS}^*}$ of the Hamiltonian \eqref{eq:HS_Ham}, the first one above the ground state that has vanishing total spin.

\noindent \textit{Thin torus limit} --- In the $\tau\to 0$ limit, $\psi_0$ and $\psi_{\ha}$ converge to the two ground states $\ket{\text{MG}}_\pm$ of the spin $\ha$ Majumdar-Ghosh chain,
\begin{equation}
    H_{MG} = \sum_j{\left(\vec S_j\cdot\vec S_{j+1} + \frac{1}{2}\vec S_j\cdot\vec S_{j+2}\right)}.
    \label{eq:MG_Ham}
\end{equation}
which are given by superpositions of the two possible dimer coverings of the chain with spin singlets.

\noindent \textit{Intermediate states} --- We have studied these idMPS as ansatz wavefunctions for the ground state and the first singlet excited state of the $J_1-J_2$ model Hamiltonian,
\begin{equation}
H(J_1, J_2) = \sum_j{\left(J_1\vec S_j\cdot\vec S_{j+1} + J_2\vec S_j\cdot\vec S_{j+2}\right)}.
\end{equation}
This is a natural choice since $J_2/J_1=0.5$ reproduces \eqref{eq:MG_Ham} while $J_2/J_1\approx0.241$ is a critical point that is close to the truncation of \eqref{eq:HS_Ham} to next-to-nearest-neighbors interactions, in the large $N$ limit. We numerically searched for the optimal values of $R$ that minimize the energy in the region $J_2\in[0.1, 0.5]$ (with $J_1$ fixed to 1) and show the fidelity per site between resulting ansatz wavefunction $\psi_0$ and the ground state in Fig.~\ref{fig:plots1} (a similar plot for $\psi_{\ha}$ can be seen in \cite{SM}). The optimal value of $R$ seems to rapidly increase after a value of $J_2$ that depends on $N$, which, for the system sizes we can probe with exact diagonalization, is far from the critical point.

\begin{figure}
    \centering
        \begin{tikzpicture}
        \def\dx{3}
        \def\dy{0.7}
            \tikzset{c/.style={insert path={circle[radius=2pt]}}}
            \filldraw (0,\dy) [c] node[left=2] {$\ket{\text{MG}, S=1}_-$} -- ++(\dx,0) [c] node[right=2] {$\ket{\text{tan}}$} ;
            \filldraw (0,0) [c] node[left=2] {$\ket{\text{MG}, S=1}_+$} -- ++(\dx,0) [c] node[right=2] {$\ket{\text{sin}}$};
            \filldraw (0,-\dy) [c] node[left=2] {$\ket{\text{AKLT}}$} -- ++(\dx,\dy) [c];
            \draw  (0, \dy+.7) node[] {$\tau = 0$} ++ (\dx,0) node[] {$\tau = i\infty$};
            \node[anchor=south] at (\dx/2, \dy-0.05) {$\small{\psi_2}$};
            \node[anchor=south] at (\dx/2, -0.05) {$\small{\psi_3}$};
            \node[anchor=south] at (\dx/2, -\dy-0.1) {$\small{\psi_4}$};
        \end{tikzpicture}
    \caption{Schematic representation of the wavefunctions arising from the SU(2)$_2$ WZW model.}
    \label{fig:schema2}
\end{figure}
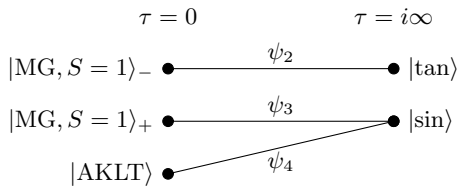

\textit{idMPS from the SU(2)$_2$ WZW model} --- The SU(2)$_2$ chiral WZW model has central charge $c=\frac{3}{2}$, and six primary fields $\phi_j^s$ grouped in three conformal towers, $\mathbf{0}$ (spin singlet), $\mathbf{\ha}$ (doublet) and $\mathbf{1}$ (triplet), with nontrivial fusion rules $\mathbf{\frac{1}{2}} \times \mathbf{\frac{1}{2}} = \mathbf{0}+\mathbf{1}$, $\mathbf{\frac{1}{2}} \times \mathbf{1} = \mathbf{\ha}$, $\mathbf{1} \times \mathbf{1} = \mathbf{0}$. It has a free field representation by three Majorana fermions $\chi^s$, giving rise to the three components of the spin triplet \cite{SM},
\begin{equation}
    \phi_1^s(z)\longmapsto\chi^s(z), \qquad s=\pm1,0.
\end{equation}
The space of conformal blocks for (even) $N$ insertions on the torus is three-dimensional, corresponding to the three allowed labellings of the fusion graph,
\begin{equation}
    \begin{tikzpicture}
    \def\dx{0.45}
    \def\dxb{0.65}
    \def\ddx{0.2}
    \def\ddxb{0.4}
    \def\dy{0.4}
    \def\ddy{0.3}
    \def\dyb{0.6}
        \draw (-\ddx, 0) -- (2*\dx+\dxb+\ddxb, 0) -- (2*\dx+\dxb+\ddxb, -\ddy) -- (-\ddx, -\ddy) -- cycle;
        \foreach \x in {0,\dx,2*\dx, 2*\dx+\dxb}{
        \draw (\x,0) --++ (0, \dy);
        \node[] at (\x, \dyb) {$_1$};}
        \node[anchor = south] at (2*\dx+0.5*\dxb, -0.05) {$_{\ldots}$};
        \node[anchor = south] at (0.5*\dx, -0.05) {$_1$};
        \node[anchor = south] at (1.5*\dx, -0.05) {$_0$};
        \node[anchor = south] at (2*\dx+\dxb+0.5*\ddxb, -0.05) {$_0$};
    \end{tikzpicture}
    \,,~
    \begin{tikzpicture}
    \def\dx{0.45}
    \def\dxb{0.65}
    \def\ddx{0.2}
    \def\ddxb{0.4}
    \def\dy{0.4}
    \def\ddy{0.3}
    \def\dyb{0.6}
        \draw (-\ddx, 0) -- (2*\dx+\dxb+\ddxb, 0) -- (2*\dx+\dxb+\ddxb, -\ddy) -- (-\ddx, -\ddy) -- cycle;
        \foreach \x in {0,\dx,2*\dx, 2*\dx+\dxb}{
        \draw (\x,0) --++ (0, \dy);
        \node[] at (\x, \dyb) {$_1$};}
        \node[anchor = south] at (2*\dx+0.5*\dxb, -0.05) {$_{\ldots}$};
        \node[anchor = south] at (0.5*\dx, -0.05) {$_0$};
        \node[anchor = south] at (1.5*\dx, -0.05) {$_1$};
        \node[anchor = south] at (2*\dx+\dxb+0.5*\ddxb, -0.05) {$_1$};
    \end{tikzpicture}
    \,,~
    \begin{tikzpicture}
    \def\dx{0.45}
    \def\dxb{0.65}
    \def\ddx{0.2}
    \def\ddxb{0.4}
    \def\dy{0.4}
    \def\ddy{0.3}
    \def\dyb{0.6}
        \draw (-\ddx, 0) -- (2*\dx+\dxb+\ddxb, 0) -- (2*\dx+\dxb+\ddxb, -\ddy) -- (-\ddx, -\ddy) -- cycle;
        \foreach \x in {0,\dx,2*\dx, 2*\dx+\dxb}{
        \draw (\x,0) --++ (0, \dy);
        \node[] at (\x, \dyb) {$_1$};}
        \node[anchor = south] at (2*\dx+0.5*\dxb, -0.05) {$_{\ldots}$};
        \node[anchor = south] at (0.5*\dx, -0.05) {$_\ha$};
        \node[anchor = south] at (1.5*\dx, -0.05) {$_\ha$};
        \node[anchor = south] at (2*\dx+\dxb+0.5*\ddxb, -0.05) {$_\ha$};
    \end{tikzpicture}\,.
    \label{eq:trees2}
\end{equation}
In the free fermion picture, a basis of the corresponding conformal blocks is given by closing the torus with periodic (P) or antiperiodic (A) boundary conditions alongside each of the cycles for all three fermions, giving rise to three different \textit{spin structures} (the PP sector actually vanishes due to the zero mode) \cite{Ginsparg:1988ui}. For each spin structure, the two-point function is given by
\begin{align}
    \langle\chi(z)\chi(0)\rangle_{\tau,\nu}&=\wp_\nu(z|\tau)
    \label{eq:2ptfn}
\end{align}
where $\wp_\nu(z|\tau)$ are the generalized Weierstrass functions, and $\nu=2,3,4$ labels the three spin structures PA, AA, and AP respectively \cite{SM}. The conformal block is then built by Wick's theorem, taking the Pfaffian of the matrix of two-body correlations, 
\begin{equation}
    \psi_{\nu}(\mb s|\tau) = \pf{C_\nu(\mb s|\tau)},\quad [C_\nu(\mb s|\tau)]_{ij} = \wp_\nu(z_i-z_j|\tau)\delta_{s_is_j}.
\end{equation}

\begin{figure*}

\centering
\begin{subfigure}{0.45\textwidth}
    \includegraphics[width=\textwidth]{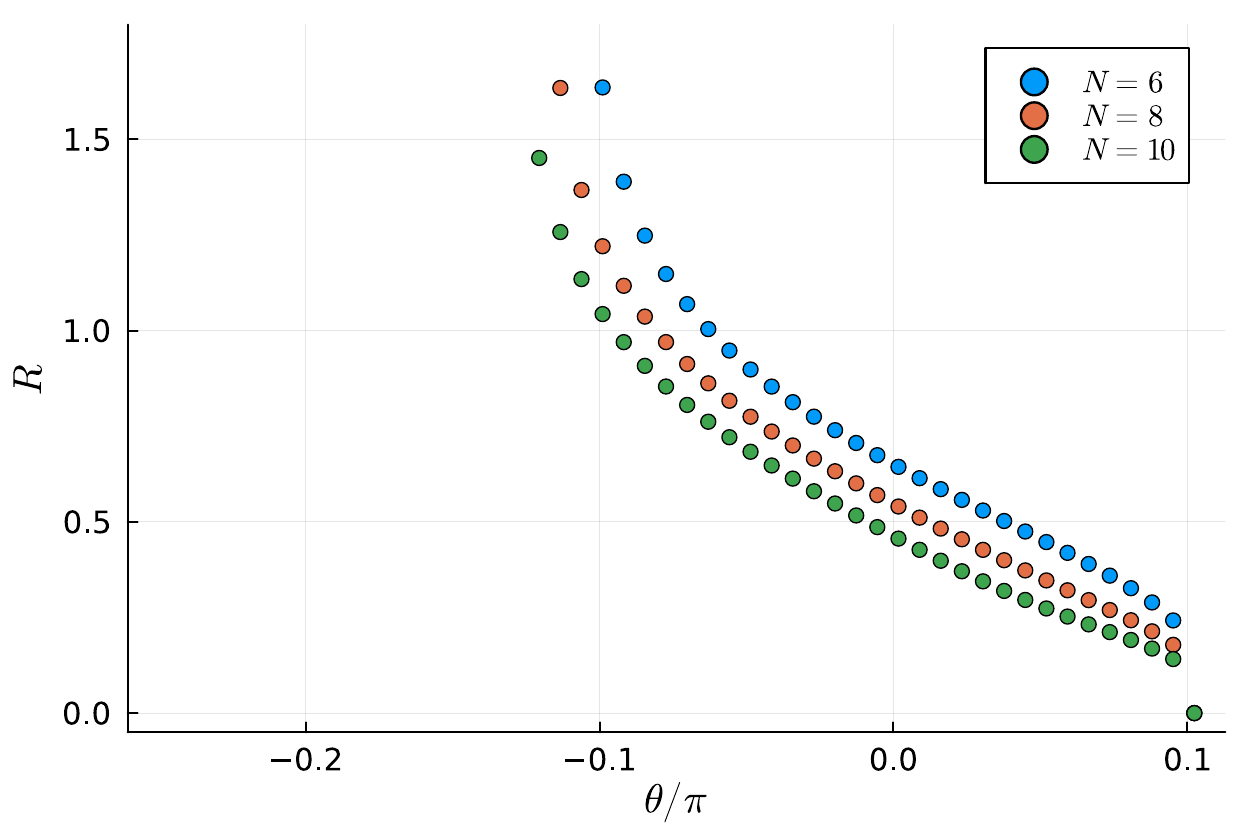}
    \caption{}
\end{subfigure}
\hfill
\begin{subfigure}{0.45\textwidth}
    \includegraphics[width=\textwidth]{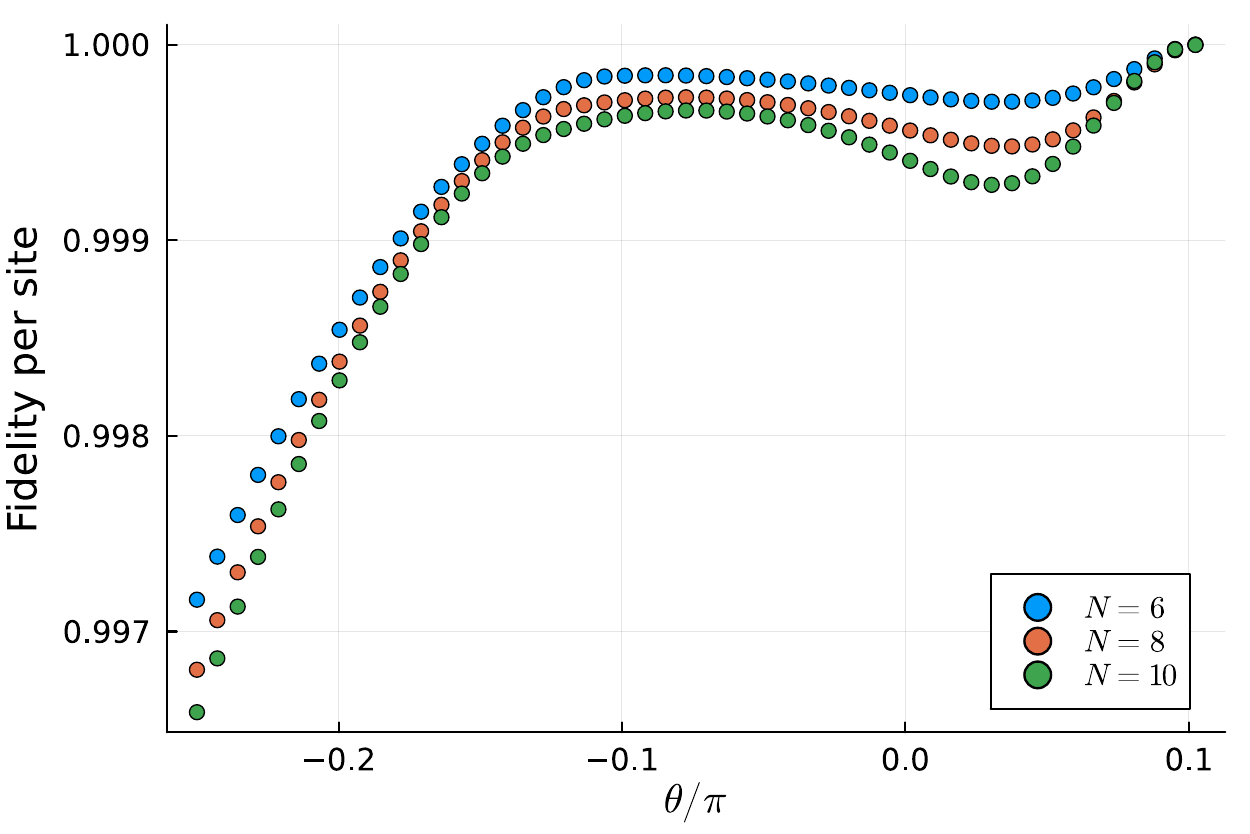}
    \caption{}
\end{subfigure}
    \caption{(a) Optimal value of $R$ for each point in the phase diagram of the quadratic-biquadratic model, labeled by $\theta$. (b) Fidelity per site of the ansatz wavefunction $\psi_4$ with the true ground state for each value of $\theta$.}
    \label{fig:plots2}
\end{figure*}

\noindent \textit{Cylinder limit} --- 
We can take the cylinder limit $\tau\to iR$ already at the level of the two-point functions \eqref{eq:2ptfn}, to find
\begin{equation}
    \lim_{\tau\to i\infty}\wp_2(z|\tau) =\dfrac{\pi}{\tan(\pi z)},~~\lim_{\tau\to i\infty}\wp_{3,4}(z|\tau) =\dfrac{\pi}{\sin(\pi z)}.
\end{equation}
Thus the two wavefunctions $\psi_3$, $\psi_4$ coincide in the cylinder limit. This is due to the number of spin structures being reduced to two with the effective disappearance of the cycle taken to infinity. We call the resulting spin states $\ket{\text{tan}}, \ket{\text{sin}}$. We can compute a (nonlocal) parent Hamiltonian for $\ket{\text{sin}}$ using the techniques from \cite{Nielsen_2011}. Numerically we then check that $\ket{\text{tan}}$ has a large overlap with an excited state of this Hamiltonian.

\noindent \textit{Thin torus limit} --- 
In the thin torus limit, $\ket{\psi_2}$ and $\ket{\psi_3}$ converge to analogous superpositions of dimer coverings as in the SU(2)$_1$, Majumdar-Ghosh case. The corresponding dimer state is $\ket{11}+\ket{00}+\ket{\text{-}1\text{-}1}$, which is not a spin singlet but rather related to the spin singlet by the local unitary transformation 
\begin{equation}
    u=\frac{1}{\sqrt{2}}\left(\begin{array}{ccc}
       -1  &  -i & 0\\
        0 & 0  & \sqrt{2} \\
       1 &  -i &  0 \\
    \end{array}\right)
    \label{eq:u}
\end{equation}
which maps between the bases of linear and circular polarizations \cite{SM}. On the other hand, $\ket{\psi_4}$ converges to the AKLT state in the same basis of circular polarizations, which corresponds to the MPS generated by the Pauli matrices. The usual presentation of the AKLT state is recovered by applying the local rotation $u$. At the level of the field theory, such a transformation can be interpreted as the bosonization of two of the three real fermions.

\noindent \textit{Intermediate states} --- 
This time we focus on the $\psi_4$ wavefunction as an ansatz for the ground state of the Haldane phase of the quadratic-biquadratic model \cite{BLBQ1, BLBQ2, BLBQ3},
\begin{equation}
    H(\theta) = \sum_j{\left(\cos\theta\,\vec S_j\cdot\vec S_{j+1} + \sin\theta\,(\vec S_j\cdot\vec S_{j+1})^2\right)}.
    \label{eq:QBQ_Ham}
\end{equation}
which is a natural choice since it includes the AKLT Hamiltonian (at $\theta=\arctan{1/3}\approx0.1\pi$) and a phase transition displaying SU(2)$_2$ criticality, the Takhtajan-Babudjian critical point at $\theta=-\pi/4$. We find the optimal value of $R$ that minimizes the energy for each Hamiltonian with $\theta\in[-\pi/4, \arctan{1/3}]$ and show the fidelity per site between the ansatz wavefunction $\psi_4$ and the ground state in Fig.~\ref{fig:plots2}. In the region close to the AKLT point, the ansatz seems to behave quite nicely, whereas the optimal value of $R$ increases rapidly at some $N$-dependent value of $\theta$ in the middle of the phase, at which point it cannot be determined accurately at these system sizes, since the wavefunction barely changes with $R$ at large values of $R$.

\textit{Outlook} --- In this Letter, we have shown that idMPS defined from CFT conformal blocks on the torus give rise to one-parameter families of spin chain wavefunctions with a finite dimensional MPS on one end, and we have given examples featuring well-known MPS arising as thin torus limits for also well-known chiral CFTs. Many questions remain open, for instance regarding the characterization of the MPS that can appear in this construction, whose tensors encode matrix elements of chiral vertex operators. Slight modifications of the AKLT example in this Letter should allow the realization of a broad class of MPS built from the algebra of Pauli matrices, such as the cluster state, and from higher Clifford algebras, such as those in \cite{Tu_2008, PhysRevB.78.094404}. It would also be worthwhile to investigate whether the connection to the virtual CFT leaves observable imprints on the many-body state, especially in its symmetries and phase classification. For MPS, these properties are known to be encoded in the structure of the local tensors, and similar results are known already for field tensor network states \cite{Gasull_2023}. Furthermore, this construction reveals an intriguing connection between the geometric data encoded in the torus’s modular parameter and the entanglement and correlation properties of the resulting many-body state. It also opens the door to potential generalizations involving more complex geometries, such as higher-genus Riemann surfaces with multi-parameter moduli spaces, which may even allow for the description of mixed states.

\textit{Acknowledgements} --- A.F.R. acknowledges support from the Alexander von Humboldt Foundation through a Postdoctoral Fellowship and from the Austrian Science Fund FWF through Grant DOI 10.55776/F71. G.S. acknowledges financial support from the Spanish MINECO grant PID2021-127726NB-I00, the CSIC Research Platform on Quantum Technologies PTI-001, the QUANTUM ENIA project Quantum Spain funded through the RTRP-Next Generation program under the framework of the Digital Spain 2026 Agenda and partial support from NSF grant PHY-2309135 to the Kavli Institute for Theoretical Physics (KITP), as well as joint sponsorship from the Fulbright Program and the Spanish Ministry of Science, Innovation and Universities. Research at MPQ is partly funded by THEQUCO as part of the Munich Quantum Valley, which is supported by the Bavarian state government with funds from the Hightech Agenda Bayern Plus. We acknowledge the ``Quantum Information Theory'' research term at the Instituto de Ciencias Matemáticas (ICMAT), the ``Quantum Information'' conference at the Centro de Ciencias de Benasque Pedro Pascual (CCBPP), and the ``Tensor Networks for Chiral Topological Phases'' programme of the International Quantum Tensor Network (IQTN) and the University of Kent, where part of this work was conducted. 

\bibliography{refs}
\clearpage
\appendix
\section{Supplemental Material}
\subsection{Special functions}
\label{app:spec_fun}
We introduce here the special functions that appear in the main text. We begin by the \textit{Jacobi theta function}, that is defined by
\begin{equation}
    \theta\genfrac[]{0pt}{0}{a}{b}(z|\tau):=\sum_{n\in \Z}e^{i\pi\tau(n+a)^2+2i\pi(z+b)(n+a)}.
\end{equation}
Usually, the following values of $a,b$ are given specific names
\begin{align}
    \theta_1:=-\theta\jfrac{\ha}{\ha},\qquad  & \theta_2:=\theta\jfrac{\ha}{0},\nonumber\\
    \theta_3:=\theta\jfrac{0}{0},\qquad  & \theta_4:=\theta\jfrac{0}{\ha}.
\end{align}
These are the only theta functions used in this work. The \textit{prime form} $E(z|\tau)$ and the \textit{generalized Weierstrass functions} $\wp_\nu(z|\tau)$ are all defined in terms of $\theta_1,\ldots,\theta_4$:
\begin{align}
    &E(z|\tau):= \dfrac{\theta_1(z|\tau)}{\theta'_1(0|\tau)},
    \nonumber\\
    &\wp_\nu(z|\tau):=
    \dfrac{1}{E(z|\tau)}\dfrac{\theta_\nu(z|\tau)}{\theta_\nu(0|\tau)},\qquad \nu=2,3,4.
\end{align}
Note that in the definition of the SU(2)$_1$ wavefunctions \eqref{eq:wfs1} we could have replaced the prime form $E(z|\tau)$ by the theta function $\theta_1(z|\tau)$ and(or) its exponent $s_is_j/2$ by $\delta_{s_i,s_j}=(1+s_is_j)/2$ without affecting the state, since the changes would amount to a global scalar factor.

\subsection{idMPS momenta}
Using the periodicity properties of the special functions above, it can be checked that the idMPS wavefunctions introduced in the main text have well-defined momentum. In particular, in the SU(2)$_1$ case we have
\begin{equation}
    T\ket{\psi_0} = e^{i\frac{N}{2}\pi}\ket{\psi_0},\quad T\ket{\psi_\ha} = e^{i\left(\frac{N}{2}+1\right)\pi}\ket{\psi_\ha},
\end{equation}
where $T$ is the translation operator that shifts the chain by one site, and the system size dependence is due to the Marshall sign. Similarly, in the SU(2)$_2$ case the three families of wavefunctions $\psi_{\nu}$ also have well-defined momenta,
\begin{equation}
T\ket{\psi_2}=-\ket{\psi_2},~~T\ket{\psi_3}=\ket{\psi_3},~~T\ket{\psi_4}=\ket{\psi_4}.
\end{equation}

\subsection{Choice of basis of conformal blocks and modular transformations}
\label{app:basis_of_blocks}
The $N$-primary correlation functions of a conformal field theory on some Riemann surface can be decomposed, thanks to the operator product expansion (OPE), into a sum of products of chiral contributions,
\begin{equation}
    \ev{\phi_1(z_1,\bar z_1)\ldots\phi_N(z_N,\bar z_N)}=\sum_{\lambda,\bar\lambda}{c_{\lambda\bar\lambda}\mc F_\lambda(\{z_i\})\overline{\mc F}(\{\bar z_i\})},
\end{equation}
where $\mc F_\lambda, \overline{\mc F}_\lambda$ are \textit{conformal blocks}. The space of conformal blocks depends on the choice of primaries and the moduli of the surface. If we fix a fusion tree (a sequential way of fusing all the primaries together given by a decomposition of the $N$-punctured surface into ``pairs of pants''), we can associate to it a basis of conformal blocks indexed by the possible labellings of this fusion tree allowed by the fusion rules. Different surface decompositions will yield different bases of conformal blocks related to each other by linear \textit{duality} transformations, under which the theory should be invariant.

\begin{figure*}
    \centering
    \includegraphics[width=0.95\textwidth]{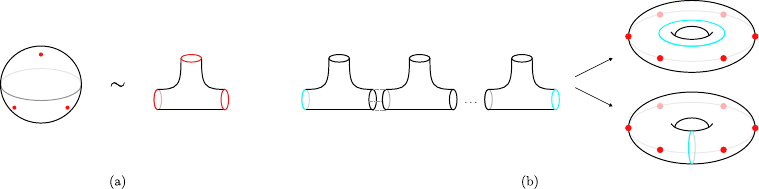}
    \caption{(a) A sphere with three punctures is topologically a pair of pants surface. (b) The torus with $N$ punctures decomposed as a sewing of $N$ pairs of pants, giving rise to two bases of conformal blocks depending on along which cycle we perform the last sewing to close the torus. We dub these bases \textit{parallel} (above) and \textit{perpendicular} (below).}
    \label{fig:confblocks}
\end{figure*}

Given the above, the choice of a Riemann surface (in our case, a torus of modular parameter $\tau=iR$ and a pattern of primary field insertions) gives rise to a potentially multidimensional subspace of idMPS wavefunctions. In this work, we use two different (albeit equivalent) decompositions of the $N$-punctured torus into pairs of pants, which just differ in what noncontractible cycle of the torus corresponds to the last sewing (see Figure \ref{fig:confblocks}). Since the $S$ modular transformation exchanges the two cycles of the torus, its linear action on the conformal blocks will effect the change of basis between the two. 

In the SU(2)$_1$ case, the two conformal blocks used in Eq.~\eqref{eq:wfs1}, computed in \cite{Dijkgraaf:1987vp}, are labeled 0 or $\ha$ according to the conformal tower we close the torus with (recall Eq.~\eqref{eq:trees1}). The configuration of sewing cycle and insertions is the one depicted at the top of Fig.~\ref{fig:confblocks} (b), where the insertions are located along a cycle parallel to the sewing condition, so we will refer to this basis as the \textit{parallel} basis for short. In the following section we will argue that the conformal blocks in the configuration at the bottom of Fig.~\ref{fig:confblocks} (b), with the insertions along a cycle that is perpendicular to the sewing cycle (which we will thus name the \textit{perpendicular} basis), naturally give rise to MPS in the thin torus limit, with MPS tensors defined in terms of matrix elements of chiral vertex operators. The MPS we obtained in the main text are then linear combinations of the former, with coefficients dictated by the modular transformation $S$ that maps the perpendicular to the parallel basis. Finite linear combinations of finite bond dimensional MPS are, of course, finite bond dimensional MPS themselves.

To compute the action of $S$ on the conformal blocks, we can use the transformation $z\mapsto w=(z-1)/\tau$, that maps the parameterization $z=a+b\tau,~a,b\in[0,1]$ of the torus of modular parameter $\tau$ to that of the torus of modular parameter $-1/\tau$. Using the following transformation properties of the special functions involved (derived from Poisson resummation \cite{dMFS}),
\begin{align}
\theta_{\nu}\left(\frac{z}{\tau}\left|-\frac{1}{\tau}\right.\right)&=(-i)^{\delta_{\nu,1}}\sqrt{-i\tau}\,e^{i\pi\frac{z^2}{\tau}}\theta_{\tilde\nu}(z|\tau)\\
E\left(\frac{z}{\tau}\left|-\frac{1}{\tau}\right.\right)&=\frac{e^{i\pi\frac{z^2}{\tau}}}{\tau}E(z|\tau)\\
\theta_3\left(\frac{z}{\tau}\left|-\frac{2}{\tau}\right.\right)&=\sqrt{-i\tau}\,\frac{e^{i\pi\frac{z^2}{\tau}}}{\sqrt 2}\left(\theta_3(z|\tau)+\theta_2(z|\tau)\right)\\
\theta_2\left(\frac{z}{\tau}\left|-\frac{2}{\tau}\right.\right)&=\sqrt{-i\tau}\,\frac{e^{i\pi\frac{z^2}{\tau}}}{\sqrt 2}\left(\theta_3(z|\tau)-\theta_2(z|\tau)\right)
\end{align}
where $\tilde\nu$ denotes the permutation $(1,2,3,4)\mapsto (1,4,3,2)$, it can be shown that the $S$ transformation maps the conformal blocks as follows,
\begin{align}
    \begin{tikzpicture}[baseline=0.18cm]
    \def\dx{0.7}
    \def\dy{0.5}
        \draw (0, \dy) -- (0,0) -- (\dx, 0) -- (\dx, \dy);
        \draw[thick, blue, decorate, decoration = {snake, segment length=1.2mm, amplitude=.2mm}] (\dx, \dy) -- (0, \dy);
        \node[blue] at (\dx/2, \dy+0.2) {{\small $0$}};
    \end{tikzpicture}
    =
    \frac{1}{\sqrt{2}}\;
    \begin{tikzpicture}[baseline=.18cm]
    \def\dx{0.7}
    \def\dy{0.5}
        \draw  (\dx, \dy) -- (0, \dy) -- (0,0) -- (\dx, 0);
        \draw[thick, blue, decorate, decoration = {snake, segment length=1.2mm, amplitude=.2mm}] (\dx, 0) -- (\dx, \dy);
        \node[blue] at (\dx+0.2, \dy/2) {{\small $0$}};
    \end{tikzpicture}
    +
    \frac{1}{\sqrt{2}}\;
    \begin{tikzpicture}[baseline=.18cm]
    \def\dx{0.7}
    \def\dy{0.5}
        \draw  (\dx, \dy) -- (0, \dy) -- (0,0) -- (\dx, 0);
        \draw[thick, blue, decorate, decoration = {snake, segment length=1.2mm, amplitude=.2mm}] (\dx, 0) -- (\dx, \dy);
        \node[blue] at (\dx+0.2, \dy/2) {{\small $\ha$}};
        \label{eq:firstS}
    \end{tikzpicture},\\
    \begin{tikzpicture}[baseline=0.18cm]
    \def\dx{0.7}
    \def\dy{0.5}
        \draw (0, \dy) -- (0,0) -- (\dx, 0) -- (\dx, \dy);
        \draw[thick, blue, decorate, decoration = {snake, segment length=1.2mm, amplitude=.2mm}] (\dx, \dy) -- (0, \dy);
        \node[blue] at (\dx/2, \dy+0.28) {{\small $\ha$}};
    \end{tikzpicture}
    =
    \frac{1}{\sqrt{2}}\;
    \begin{tikzpicture}[baseline=.18cm]
    \def\dx{0.7}
    \def\dy{0.5}
        \draw  (\dx, \dy) -- (0, \dy) -- (0,0) -- (\dx, 0);
        \draw[thick, blue, decorate, decoration = {snake, segment length=1.2mm, amplitude=.2mm}] (\dx, 0) -- (\dx, \dy);
        \node[blue] at (\dx+0.2, \dy/2) {{\small $0$}};
    \end{tikzpicture}
    -
    \frac{1}{\sqrt{2}}\;
    \begin{tikzpicture}[baseline=.18cm]
    \def\dx{0.7}
    \def\dy{0.5}
        \draw  (\dx, \dy) -- (0, \dy) -- (0,0) -- (\dx, 0);
        \draw[thick, blue, decorate, decoration = {snake, segment length=1.2mm, amplitude=.2mm}] (\dx, 0) -- (\dx, \dy);
        \node[blue] at (\dx+0.2, \dy/2) {{\small $\ha$}};
    \end{tikzpicture}.
\end{align}
In the case of SU(2)$_2$ represented by three free fermions, our three conformal blocks defined by different spin structures provide yet another basis of the space of conformal blocks. It is related to the bases labeled by the conformal towers by the following linear combinations \cite{Ginsparg:1988ui},
\begin{align}
    \begin{tikzpicture}[baseline=.18cm]
    \def\dx{0.7}
    \def\dy{0.5}
        \draw  (\dx, \dy) -- (0, \dy) -- (0,0) -- (\dx, 0)-- (\dx, \dy);
        \node[anchor=west] at (\dx, \dy/2) {{A}};
        \node[anchor=north] at (\dx/2, 0) {{P}};
    \end{tikzpicture}
    &=
    \begin{tikzpicture}[baseline=.18cm]
    \def\dx{0.7}
    \def\dy{0.5}
        \draw  (\dx, \dy) -- (0, \dy) -- (0,0) -- (\dx, 0);
        \draw[thick, blue, decorate, decoration = {snake, segment length=1.2mm, amplitude=.2mm}] (\dx, 0) -- (\dx, \dy);
        \node[blue] at (\dx+0.2, \dy/2) {{\small $0$}};
    \end{tikzpicture}
    -
\begin{tikzpicture}[baseline=.18cm]
    \def\dx{0.7}
    \def\dy{0.5}
        \draw  (\dx, \dy) -- (0, \dy) -- (0,0) -- (\dx, 0);
        \draw[thick, blue, decorate, decoration = {snake, segment length=1.2mm, amplitude=.2mm}] (\dx, 0) -- (\dx, \dy);
        \node[blue] at (\dx+0.2, \dy/2) {{\small $1$}};
    \end{tikzpicture},\\
   \begin{tikzpicture}[baseline=.18cm]
    \def\dx{0.7}
    \def\dy{0.5}
        \draw  (\dx, \dy) -- (0, \dy) -- (0,0) -- (\dx, 0)-- (\dx, \dy);
        \node[anchor=west] at (\dx, \dy/2) {{A}};
        \node[anchor=north] at (\dx/2, 0) {{A}};
    \end{tikzpicture}
    &=
    \begin{tikzpicture}[baseline=.18cm]
    \def\dx{0.7}
    \def\dy{0.5}
        \draw  (\dx, \dy) -- (0, \dy) -- (0,0) -- (\dx, 0);
        \draw[thick, blue, decorate, decoration = {snake, segment length=1.2mm, amplitude=.2mm}] (\dx, 0) -- (\dx, \dy);
        \node[blue] at (\dx+0.2, \dy/2) {{\small $0$}};
    \end{tikzpicture}
    +
\begin{tikzpicture}[baseline=.18cm]
    \def\dx{0.7}
    \def\dy{0.5}
        \draw  (\dx, \dy) -- (0, \dy) -- (0,0) -- (\dx, 0);
        \draw[thick, blue, decorate, decoration = {snake, segment length=1.2mm, amplitude=.2mm}] (\dx, 0) -- (\dx, \dy);
        \node[blue] at (\dx+0.2, \dy/2) {{\small $1$}};
    \end{tikzpicture},\label{eq:lastS}\\
\begin{tikzpicture}[baseline=.18cm]
    \def\dx{0.7}
    \def\dy{0.5}
        \draw  (\dx, \dy) -- (0, \dy) -- (0,0) -- (\dx, 0)-- (\dx, \dy);
        \node[anchor=west] at (\dx, \dy/2) {{P}};
        \node[anchor=north] at (\dx/2, 0) {{A}};
    \end{tikzpicture}
    &=\sqrt{2}\;
    \begin{tikzpicture}[baseline=.18cm]
    \def\dx{0.7}
    \def\dy{0.5}
        \draw  (\dx, \dy) -- (0, \dy) -- (0,0) -- (\dx, 0);
        \draw[thick, blue, decorate, decoration = {snake, segment length=1.2mm, amplitude=.2mm}] (\dx, 0) -- (\dx, \dy);
        \node[blue] at (\dx+0.2, \dy/2) {{\small $\ha$}};
    \end{tikzpicture}.
\end{align}

\subsection{Thin torus limit argument}
\label{app:thin_limit}
In this section we present the argument why the thin torus limit of our wavefunctions yields finite dimensional matrix product states. The argument is based on the decomposition of the torus conformal blocks as traces of chiral vertex operators. This procedure is accompanied by a twin geometrical picture consisting on the sewing of punctured Riemann surfaces 

Let us go through the geometric picture first. We will construct a conformal block by sewing $n$ identical copies of the Riemann sphere with three punctures at $z=0, 1, \infty$, i.e.~the punctured Riemann surface $\CP^1\backslash\{0,1,\infty\}$. The sewing proceeds by selecting local coordinates $z, w$ around the two punctures to be sewn (such that the punctures sit at $z=0$ and $w=0$ respectively), eliminating small disks around each of them and identifying the annuli outside said disks via the condition $zw = q$, where $q$ is a complex number that will parameterize the family of punctured Riemann surfaces that may arise from the sewing operation (i.e., $q$ is a coordinate of the \textit{moduli space} for the corresponding genus and number of punctures). Since all of our surfaces are identical Riemann spheres, we can cover each of them by two patches of local coordinates $z_i, w_i$ with transition function $w_i = 1/z_i$. We will then consistently glue them at the punctures $z_i=0$ and $z_i = \infty$ by identifying the local coordinates as $z_{i+1}w_i=q$, using the same sewing parameter $q$ to maintain the symmetry. For later convenience we define 
\begin{equation}
    q \equiv \exp{\left(-\frac{2\pi }{nR}\right)},\qquad R>0,
    \label{eq:def_q}
\end{equation}
so $q<1$.

The first sewing operation consists of gluing the $z_1=\infty$ puncture of sphere 1 with the $z_2=0$ puncture of sphere 2, and results in a Riemann sphere with four punctures, which can be described in terms of the $z_2$ coordinate as sitting at positions $z_2 = 0, q, 1, \infty$. We iterate the procedure with the remaining $n-2$ spheres and reach a Riemann sphere with $n+2$ punctures at $z_n = 0, q^{n-1}, q^{n-2},\ldots, q, 1, \infty$. The last gluing step now operates on two punctures on the same surface, namely at $z_n=0$ and $z_n=\infty$. Note however, that the sewing identification involves the original coordinate $z_1$, which has been rescaled by $q$ $n-1$ times. Thus the identification is $z_1w_n = q\implies z_n w_n/q^{n-1} = q\implies z_n \sim q^n z_n$. The resulting Riemann surface is thus periodic in the radial coordinate, and we can see that it corresponds to a torus with punctures by applying the conformal change of coordinates,
\begin{equation}
    z_n = \exp{\left(\frac{2\pi t}{R}\right)}.
    \label{eq:z2t}
\end{equation}
Indeed, in the new coordinate $t$ our Riemann surface corresponds to a rectangle $t\in [0,1]\times[0, iR]$ with periodic boundary conditions in both directions, which is to say a torus of modular parameter $\tau = iR$. The coordinates of the $n$ remaining punctures are $t=1/n, \ldots, (n-1)/n, 1$, and the last sewing can be taken to be performed along the circumference $|z|=1$, corresponding to the segment $t\in i[0,1]$. Thus, by sewing spheres we have recomposed the $N$-punctured torus, and the particular choice of coordinates and mappings in this argument corresponds to the lower path in Figure \ref{fig:confblocks}, i.e. the perpendicular basis.

Now we switch to the algebraic picture of this same procedure. A three-punctured sphere with punctures at $\infty,z,0$ labeled by chiral algebra representations $\hilb_i,\hilb_j,\hilb_k$ respectively is associated to a \textit{chiral vertex operator} \cite{Tsuchiya:1987rv, MOORE1988451},
\begin{equation}
    \begin{tikzpicture}[baseline=8]
    \def\dx{0.5}
    \def\dy{0.6}
        \draw (-\dx, 0) -- (-\dx/2, 0);
        \draw[<-] (-\dx/2, 0) -- (0, 0);
        \draw (0, 0) -- (\dx/2, 0);
        \draw [<-](\dx/2, 0) -- (\dx, 0);
        \draw (0, 0) -- (0, \dy/2);
        \draw[<-] (0, \dy/2) -- (0, \dy);
        \node[anchor = east] at (-\dx, 0) {$i$};
        \node[anchor = south] at (0, \dy) {$(j,\beta)$};
        \node[anchor = west] at (\dx, 0) {$k$};
    \end{tikzpicture}\qquad V^i_{jk}(z, \beta):\hilb_k\rightarrow\hilb_i,
\end{equation}
where $\beta\in \hilb_j$ is a state in the $j$-th representation. Let $\hilb^0_i$ be the subspace of $\hilb_i$ spanned by the states with lowest $L_0$ eigenvalue, denoted $\Delta_i$. The matrix elements of the vertex operators on these subspaces correspond to three-point functions of the corresponding chiral primaries,
\begin{equation}
    \langle\gamma|V^i_{jk}(z, \beta)\ket{\alpha} = t(\alpha\otimes\beta\otimes\gamma)z^{-(\Delta_j+\Delta_k-\Delta_i)}
    \label{eq:cvo}
\end{equation}
where $t:\left(\hilb^0_i\right)^*\otimes\hilb^0_j\otimes\hilb^0_k\rightarrow\C$ is a coefficient tensor and $\alpha\in\left(\hilb^0_i\right)^*, \beta\in\hilb^0_j, \gamma\in\hilb^0_k$. 
Sewing two spheres (along punctures with matching labels) as we did above corresponds to composing chiral vertex operators, with an intermediate dilation, generated by $L_0$ due to the rescaling by the parameter $q$. Thus, the full torus conformal block obtained after sewing the $n$ spheres is given by
\begin{equation}
    \Psi_{\{s_i\}}\propto\tr{\left(q^{L_0}V_{I_{n}}(1, s_n)\ldots q^{L_0}V_{I_{2}}(1, s_2)q^{L_0}V_{I_{1}}(1, s_1)\right)},
\end{equation}
where the indices $I_m\equiv(i_m, j_m, k_m)$ refer to the choice of module labels for each sphere's punctures, we have $k_m=i_{m-1}, k_1=i_n$ and the trace is taken in $\hilb_{k(1)}$. These choices constitute the fusion path characterizing the resulting conformal block. Now, the thin torus limit corresponds to $R\to 0$, or $q\to 0$ (see \eqref{eq:def_q}), so that the operator insertions $q^{L_0}$ effectively project on the subspaces $\hilb^0_k$ of lowest conformal weight, as all other matrix elements are exponentially suppressed. Thus we replace the chiral vertex operators $V$ by their restrictions $V^0$ to said subspaces: note that these are now finite dimensional. Omitting the dependence on positions (as they are fixed), we find that up to global factors absorbed in the normalization, 
\begin{equation}
    \Psi_{\{s_i\}}\propto\tr{\left(V^0_{I_{n}}(s_n)\ldots V^0_{I_{2}}(s_2)V^0_{I_{1}}(s_1)\right)}.
\end{equation}
This expression is precisely the amplitude of a matrix product state wavefunction, resulting from the contraction of three-legged tensors which actually correspond to the $t$ tensors introduced in \eqref{eq:cvo}. We refrain in the present work from making this argument into a mathematical proof, which would require bounding the error incurred when projecting onto the low conformal weight subspaces, and are instead satisfied with checking numerically that the thin torus limits of known conformal blocks are indeed as expected from this reasoning. To that end, let's review which MPS tensors should come up in our examples. Since we are looking at the restriction of the chiral vertex operators to the subspaces spanned by the primary multiplets, we only need the first coefficient in the OPE of a pair of primary fields, which for SU(2)$_k$ is given by the Clebsch-Gordan coefficients that govern tensor products of SU(2) irreps \cite{Recknagel_Schomerus_2013}. In the SU(2)$_1$ case, the relevant chiral vertex operators allowed by the fusion rules are of the form
\begin{equation}
    \begin{tikzpicture}[baseline=0]
    \def\dx{0.5}
    \def\dy{0.6}
        \draw (-\dx, 0) -- (-\dx/2, 0);
        \draw[<-] (-\dx/2, 0) -- (0, 0);
        \draw (0, 0) -- (\dx/2, 0);
        \draw [<-](\dx/2, 0) -- (\dx, 0);
        \draw (0, 0) -- (0, \dy/2);
        \draw[<-] (0, \dy/2) -- (0, \dy);
        \node[anchor = east] at (-\dx, 0) {$\ha, s_1$};
        \node[anchor = south] at (0, \dy) {$\ha, s_2$};
        \node[anchor = west] at (\dx, 0) {$0$};
    \end{tikzpicture}
    \propto \delta_{s_1,s_2},
\end{equation}
and
\begin{equation}
    \begin{tikzpicture}[baseline=0]
    \def\dx{0.5}
    \def\dy{0.6}
        \draw (-\dx, 0) -- (-\dx/2, 0);
        \draw[<-] (-\dx/2, 0) -- (0, 0);
        \draw (0, 0) -- (\dx/2, 0);
        \draw [<-](\dx/2, 0) -- (\dx, 0);
        \draw (0, 0) -- (0, \dy/2);
        \draw[<-] (0, \dy/2) -- (0, \dy);
        \node[anchor = east] at (-\dx, 0) {$0$};
        \node[anchor = south] at (0, \dy) {$\ha, s_1$};
        \node[anchor = west] at (\dx, 0) {$\ha, s_2$};
    \end{tikzpicture}
    \propto (-1)^{s_1-\ha}\delta_{s_1,-s_2},
\end{equation}
for $s_1,s_2\in\left\{\pm\ha\right\}$. In the SU(2)$_2$ case, we have the analogous
\begin{equation}
    \begin{tikzpicture}[baseline=0]
    \def\dx{0.5}
    \def\dy{0.6}
        \draw (-\dx, 0) -- (-\dx/2, 0);
        \draw[<-] (-\dx/2, 0) -- (0, 0);
        \draw (0, 0) -- (\dx/2, 0);
        \draw [<-](\dx/2, 0) -- (\dx, 0);
        \draw (0, 0) -- (0, \dy/2);
        \draw[<-] (0, \dy/2) -- (0, \dy);
        \node[anchor = east] at (-\dx, 0) {$1, s_1$};
        \node[anchor = south] at (0, \dy) {$1, s_2$};
        \node[anchor = west] at (\dx, 0) {$0$};
    \end{tikzpicture}
    \propto \delta_{s_1,s_2},
\end{equation}
and
\begin{equation}
    \begin{tikzpicture}[baseline=0]
    \def\dx{0.5}
    \def\dy{0.6}
        \draw (-\dx, 0) -- (-\dx/2, 0);
        \draw[<-] (-\dx/2, 0) -- (0, 0);
        \draw (0, 0) -- (\dx/2, 0);
        \draw [<-](\dx/2, 0) -- (\dx, 0);
        \draw (0, 0) -- (0, \dy/2);
        \draw[<-] (0, \dy/2) -- (0, \dy);
        \node[anchor = west] at (\dx, 0) {$1, s_1$};
        \node[anchor = south] at (0, \dy) {$1, s_2$};
        \node[anchor = east] at (-\dx, 0) {$0$};
    \end{tikzpicture}
    \propto (-1)^{s_1-1}\delta_{s_1, -s_2},
\end{equation}
where now $s_1,s_2\in\{1,0,-1\}$. In both cases, clearly one of the tensors is just an identity map, while the other corresponds to the SU(2) singlet state, so that the alternative contraction of both tensors gives rise to a dimer covering of the chain by singlets, and the Majumdar-Ghosh ground states arise due to the linear combinations of the $S$ modular transformations \eqref{eq:firstS}-\eqref{eq:lastS}. For SU(2)$_2$, however, we also have the genuinely three-legged tensor 
\begin{equation}
    \begin{tikzpicture}[baseline=0]
    \def\dx{0.5}
    \def\dy{0.6}
        \draw (-\dx, 0) -- (-\dx/2, 0);
        \draw[<-] (-\dx/2, 0) -- (0, 0);
        \draw (0, 0) -- (\dx/2, 0);
        \draw [<-](\dx/2, 0) -- (\dx, 0);
        \draw (0, 0) -- (0, \dy/2);
        \draw[<-] (0, \dy/2) -- (0, \dy);
        \node[anchor = east] at (-\dx, 0) {$\ha, s_1$};
        \node[anchor = south] at (0, \dy) {$1, s_2$};
        \node[anchor = west] at (\dx, 0) {$\ha,s_3$};
    \end{tikzpicture}
    \propto \begin{cases}
        -\delta_{s_1,\ha}\delta_{s_3,-\ha} & s_2=1,\\\\
        \dfrac{(-1)^{s_1-\frac{1}{2}}}{\sqrt{2}}\delta_{s_1,s_3} &  s_2=0,\\\\
        \delta_{s_1,-\ha}\delta_{s_3,\ha} & s_2=-1,
    \end{cases}
    \label{eq:PaulimatMPS}
\end{equation}
where $s_1,s_3\in\left\{\pm\ha\right\}$ and $s_2\in\{1,0,-1\}$. This is a well-known MPS tensor representing the AKLT state, and, under the change of basis induced by \eqref{eq:u} (see also \eqref{eq:cob}) on the physical index, it gives the Pauli matrix tensor ${A^a\propto\sigma^a}$, as we obtained in the main text.

\subsection{Free field representations}
Here we review in a bit more detail the free field representations that we use to compute the conformal blocks of the WZW CFTs in this work. Recall that a WZW model is characterized by a Kac-Moody symmetry algebra, generated by a set of currents $J^a(z)$ with OPE
\begin{equation}
    J^a(z)J^b(w)\sim \dfrac{k\,\kappa_{a,b}}{(z-w)^2}+\dfrac{i f_{abc}J^c(w)}{z-w},
    \label{eq:KMalg}
\end{equation}
where $k\in\Z$ is the level and as usual nonsingular terms are omitted. $\kappa_{a,b}$ and $f_{abc}$ contain information about the underlying Lie algebra $\mathfrak{g}$. In particular, given a set of generators $\{t^a\}_a$ of $\mathfrak{g}$, they are respectively defined by
\begin{equation}
    \kappa_{ab}:=\kappa(t^a,t^b),\qquad [t^a, t^b]=if_{abc}t^c,
\end{equation}
where $\kappa$ is the Killing form with the usual normalization,
\begin{equation}
    \kappa(x, y):=\frac{1}{2h^\vee}{\tr{(\ad_x\ad_y)}},
\end{equation}
and $f_{abc}$ are therefore the structure constants of $\mathfrak{g}$. The OPE \eqref{eq:KMalg} codifies the algebraic relations of the modes of these currents, which generate an affinization of the Lie algebra $\mathfrak{g}$. Primary fields then come in multiplets associated to a representation $j$ of $\mathfrak{g}$, so that they satisfy the OPE
\begin{equation}
    J^a(z)\,\varphi^s_{j}(w)\sim\dfrac{(t^a_{(r)})_{ss'}\,\varphi^{s'}_{j}(w)}{z-w}
    \label{eq:KMprim}
\end{equation}
(some references include here a conventional minus sign). 

Let's know look at the specific examples in this paper. Considering the spin basis of $\mathfrak{su}(2)$, $S^i:=\sigma^i/2$, and $S^{\pm}:=S_x\pm iS_y$ we have following nontrivial $\kappa_{ab}$ and $f_{abc}$,
\begin{align}
    &[S^z, S^\pm]=\pm S^{\pm},\quad &[S^+,S^-]=2S^z,\nonumber\\&\kappa(S^z, S^z)=1/2, \quad &\kappa(S^+, S^-)=1.
\end{align}
It can then be checked that, given a free scalar field $\varphi$ with OPE
\begin{equation}
    \varphi(z)\varphi(w)\sim-\ln{(z-w)}
\end{equation}
the currents
\begin{equation}
    J^z:=i\partial\varphi,\qquad J^{\pm}:= e^{\pm i\sqrt{2}\varphi},
\end{equation}
satisfy the OPE \eqref{eq:KMalg} with $k=1$ \cite{dMFS}. Also, the doublet $\{e^{is\frac{\varphi(z)}{\sqrt{2}}}\}, s=\pm1$ satisfies the OPE \eqref{eq:KMprim} in the fundamental representation $t^a=S^a$. There is however an additional subtlety: to compute the conformal blocks of the SU(2)$_1$ WZW model we can't just simply replace each primary field by the corresponding vertex operator $e^{is\frac{\varphi(z)}{\sqrt{2}}}$, since this does not result in an SU(2) invariant object. For instance, the two-point function
\begin{equation}
    \ev{e^{is_1\frac{\varphi(z)}{\sqrt{2}}}e^{is_2\frac{\varphi(z)}{\sqrt{2}}}}\propto \delta_{s_1+s_2,0}
\end{equation}
is not a singlet in the spin indices. To fix that, we include the Marshall sign \eqref{eq:MS}, which then results in a two-point function
\begin{equation}
    \ev{e^{is_1\frac{\varphi(z)}{\sqrt{2}}}e^{is_2\frac{\varphi(z)}{\sqrt{2}}}}\propto s_1\delta_{s_1+s_2,0}
\end{equation}
which is a singlet. The Marshall sign can be interpreted as resulting from replacing the field insertions at odd positions by those associated to the conjugate representation $\mathbf{\ha^*}\cong\mathbf{\ha}$. From the many-body point of view, it amounts by acting on the wavefunction with Pauli $Z$ operators on odd sites in the chain.

Moving on to SU(2)$_2$, we see that using three real fermions $\{\chi^a\}_{a=x,y,z}$, with OPE
\begin{equation}
    \chi^a(z)\chi^b(w)\sim\dfrac{\delta_{ab}}{z-w}, 
\end{equation}
the currents
\begin{equation}
    J^a:=-i\varepsilon_{abc}\chi_b\chi_c,\qquad J^{\pm}:= J^x\pm iJ^y,
\end{equation}
satisfy the OPE \eqref{eq:KMalg} with $k=2$, and the field triplet
\begin{equation}
    \phi^1_1:=-\dfrac{\chi^x+i\chi^y}{\sqrt{2}},\quad \phi^0_1:=\chi^z, \quad\phi^{\text{-}1}_1=\dfrac{\chi^x-i\chi^y}{\sqrt{2}},
    \label{eq:cob}
\end{equation}
fulfills the OPE \eqref{eq:KMprim} in the spin 1 representation
\begin{equation}
    t^z=\sum_{s=-1}^1{s\ketbra{s}{s}},\quad t^+=\sqrt{2}\sum_{s=-1}^0{\ketbra{s+1}{s}},\quad t^-=(t^+)^\dagger.
\end{equation}
The basis of fermions $\chi^a$ used to define the insertions in the main text is thus related to the usual spin 1 basis by \eqref{eq:cob}, which is the change of basis effected by $u$ in \eqref{eq:u}. It can be checked that these representation does not require a correction such as the Marshall sign in the previous case, since the resulting conformal blocks are already SU(2) invariant, assuming that we pick the representation of SU(2) that matches the basis we work in.

\subsection{Excited state of the Haldane-Shastry model}
\label{app:HS_exc}

\begin{figure}
    \centering
    \includegraphics[width=\linewidth]{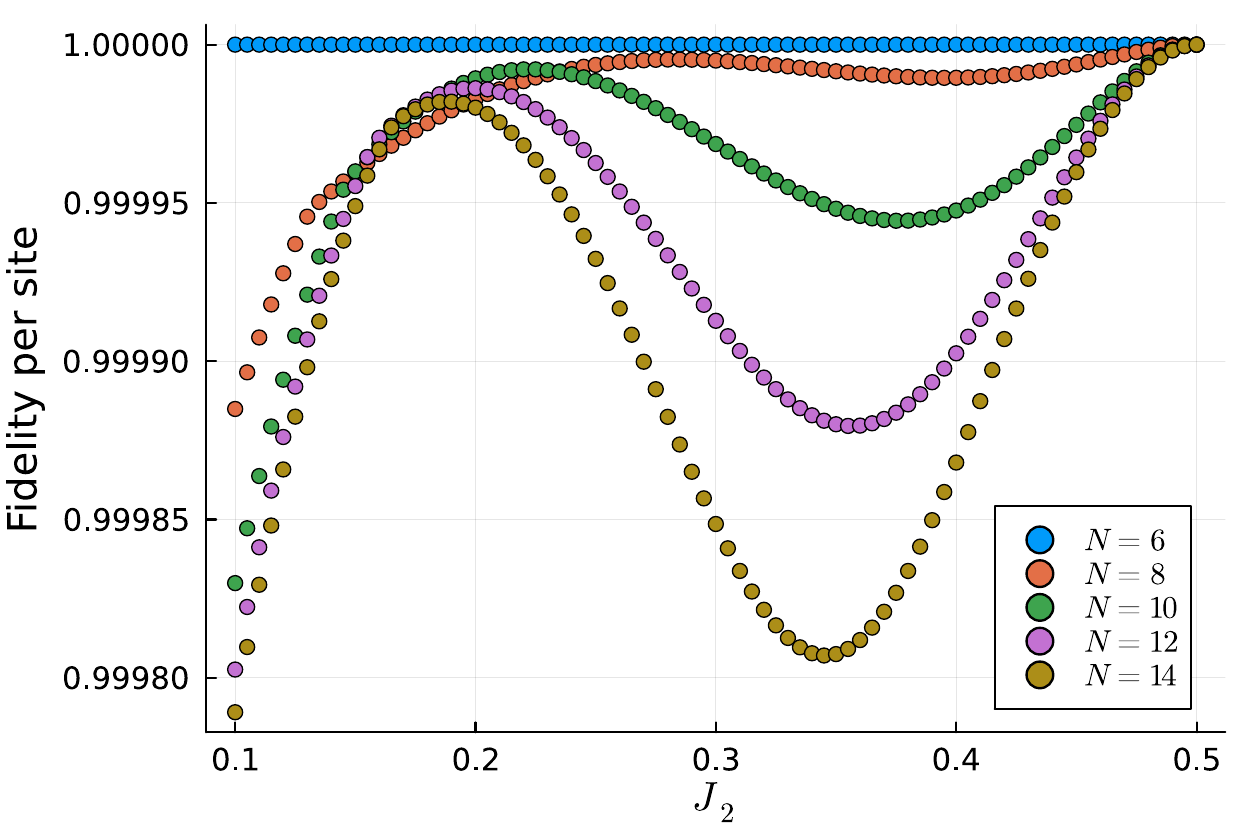}
    \caption{Fidelity between the first spin singlet excited state of the $J_1-J_2$ model and the $\psi_\ha$ idMPS at the optimal radius $R$  obtained from $\psi_0$ and the ground state.}
    \label{fig:fid12}
\end{figure}

Here we will show that the $\psi_\ha$ idMPS wavefunction built from the corresponding SU(2)$_1$ conformal block \eqref{eq:wfs1} is, in the $R\to \infty$ limit, an excited state of the Haldane-Shastry Hamiltonian, whose ground state is given by $\psi_0$. In what follows, hence, we only refer to $R=\infty$ idMPS. 

For our result we will need the following representation of $\psi_\ha$, introduced in \cite{EdgeStates}. Consider the $\psi_0$ idMPS for $N+2$ insertions at positions $z_0, z_1,\ldots,z_{N+1}$ in the limit where two insertions move towards the two extremes of the cylinder, $z_0\to-i\infty, z_{N+1}\to i\infty$. If we project the associated spins onto the singlet state $\ket{\text{sgl}}:=\frac{1}{\sqrt{2}}(\ket{01}-\ket{10})$, the resulting wavefunction of $N$ spins is precisely the $\psi_\ha$ idMPS,
\begin{align}
    &\ket{\psi_{\ha}(z_1,\ldots,z_N)}\propto\nonumber\\
    &~~~~\propto\lim_{\substack{z_0\to -i\infty\\z_{N+1}\to i\infty}} \bra{\text{sgl}}_{0,N+1}\ket{\psi_0(z_0,z_1,\ldots, z_N,z_{N+1})},
\end{align}
Our proof is then based on the formalism of \cite{Nielsen_2011}, where it is shown how to build a positive semidefinite parent Hamiltonian for an idMPS, using null vectors of the CFT. In particular, the parent Hamiltonian for the SU(2)$_1$ case reads
\begin{equation}
    H =-\sum_{\substack{i,j\\i< j}}^N\left[\dfrac{w_{ij}^2}{4}+\frac{1}{3}\left(w_{ij}^2+\sum_{\substack{k\\k\neq i,j}}{w_{ki}w_{kj}}\right)\vec S_i\cdot\vec S_j\right],
    \label{eq:pH}
\end{equation}
where $w_{jk}:=i\cot{\pi(z_j-z_k)}$ for coordinates in the cylinder, $z\in[0,1]\times i\R$, and $\vec S_i:=\ha(\sigma^x_i,\sigma^y_i,\sigma^z_i) $ is the spin operator at position $i$ \footnote{Note that in \cite{Nielsen_2011} the expressions are given in terms of complex plane coordinates, related to the cylinder coordinates as usual by an exponential map, $x+it\to e^{-2\pi i(x+it)}$.}. 

For $N+2$ insertions at $z_0\to 0$, $z_j = \frac{j}{N}, j=1,\ldots N$ and $z_{N+1}\to\infty$, the Hamiltonian \eqref{eq:pH} reads
\begin{align}
    H = H_{\text{HS}}&-\frac{N+3}{6}\vec{S}^2-\frac{2}{3}\vec S\cdot(\vec S_0+\vec S_{N+1})\nonumber\\&+\frac{N-1}{6}(\vec S_0+\vec S_{N+1})^2 -E_{\text{exc}},
\end{align}
where $H_{\text{HS}}$ is the Haldane-Shastry Hamiltonian \eqref{eq:HS_Ham} on the $z_1,\ldots,z_N$ insertions, $\vec S:=\sum_{j=1}^N{\vec S_j}$ is the total spin operator of these $N$ insertions and
\begin{equation}
    E_{\text{exc}}:=E_0+\dfrac{N}{2},\qquad E_0:=-\dfrac{N^3+5N}{24},
\end{equation}
so that $E_0$ is the ground state energy of $H_{\text{HS}}$. Then
\begin{equation}
    \bra{\text{sgl}}_{0,N+1}H=\left(H_{\text{HS}}-\frac{N+3}{6}\vec{S}^2-E_{\text{exc}}\right)\bra{\text{sgl}}_{0,N+1},
\end{equation}
and we have
\begin{align}
    H\ket{\psi_0(z_0,z_1,\ldots, z_N,z_{N+1})}=0,\\\implies\bra{\text{sgl}}_{0, N+1}H\ket{\psi_0(z_0,z_1,\ldots, z_N,z_{N+1})}=0,\\
    \left(H_{\text{HS}}-\frac{N+3}{6}\vec{S}^2-E_{\text{exc}}\right)\ket{\psi_\ha(z_1,\ldots, z_N)}=0,\\
    \left(H_{\text{HS}}-E_{\text{exc}}\right)\ket{\psi_\ha}=0,
\end{align}
since $\ket{\psi_\ha}$ is a spin singlet. Therefore, the $\psi_\ha$ idMPS is an excited state of the Haldane-Shastry Hamiltonian with energy $E_{\text{exc}}$. In the parametrization of \cite{PhysRevLett.60.635} of the eigenenergies in terms of sets of integers $\{m_j\}\subset\{1,\ldots,N-1\}$ (corresponding to the pseudomomenta, up to a factor of $2\pi/N$), the ground state corresponds to the sequence of odd numbers $(1,3,\ldots, N-1)$ and the excited state to the sequence of even numbers $(2, 4, \ldots, N-2)$. 

In Fig.~\ref{fig:fid12} we show the overlap between the first spin singlet excited state of the $J_1$-$J_2$ model and the $\psi_\ha$ idMPS for small system sizes, analogously to what we did for the ground state in Fig.~\ref{fig:plots1}(b).

\end{document}